\pdfoutput=1
\documentclass[prd,aps,preprint,amsmath,nofootinbib,amssymb,eqsecnum,showkeys,tightenlines,superscriptaddress]{revtex4-1}
\usepackage{slashed}
\usepackage{epsfig,latexsym,cancel,amssymb,amsmath,verbatim,mathrsfs}
\usepackage{color}

\usepackage{subfigure}
\usepackage{graphicx}

\begin{document}

\title{Constraints on Pseudo-Nambu-Goldstone dark matter from direct detection experiment and neutron star reheating temperature}

\author{Yu-Pan Zeng}
\email[E-mail: ]{zengyp8@mail2.sysu.edu.cn}
\affiliation{School of Physics, Sun Yat-sen University, Guangzhou, Guangdong 510275, China}

\author{Xiang Xiao}
\email[E-mail: ]{xiaox93@mail.sysu.edu.cn}
\affiliation{School of Physics, Sun Yat-sen University, Guangzhou, Guangdong 510275, China}

\author{Wei Wang}
\email[E-mail: ]{wangw223@mail.sysu.edu.cn}
\affiliation{School of Physics, Sun Yat-sen University, Guangzhou, Guangdong 510275, China}
\affiliation{Sino-French Institute of Nuclear Engineering and Technology, Sun Yat-sen University, Zhuhai, Guangdong 519082,
China}

\date{\today}

\begin{abstract}

    Pseudo-Nambu-Goldstone dark matter can interact with Standard Model fermion through the Higgs portal, and some models can emerge a cancellation mechanism that helps to escape the stringent constraints from direct detection experiments. In this paper we explore new constraints of these cancellation models on parameter space of non-zero momentum transfer, from both the current direct detection experiment and the neutron star temperature via the dark matter reheating mechanism. 
    
\end{abstract}


\maketitle
\section{Introduction} 
The nature of dark matter (DM) has been an unsolved puzzle for decades. As an appealing DM candidate, the Weakly Interacting Massive Particle (WIMP) has been pursued by plenty of direct detection experiments with ever-improving sensitivities, while the null results have put the WIMP model under pressure. Various Pseudo-Nambu-Goldstone dark matter (PNGDM) models~\cite{Gross2017,Alanne:2018zjm,Karamitros:2019ewv,Jiang:2019soj} have been proposed to relieve this tension. By introducing extra scalars with softly breaking symmetries and a mixing between the Higgs boson, PNGDM models enable a cancellation between multiple mediators, leading to the vanish of direct detection in the zero momentum transfer limit. Various studies have been performed on the framework of PNGDM models, for example, the UV completion~\cite{Abe:2020iph,Okada:2020zxo,Abe:2021byq,Okada:2021qmi}, direct detection from loop corrections ~\cite{Azevedo:2018exj,Ishiwata:2018sdi,Glaus:2020ihj,Alanne:2018zjm,Arina:2019tib}, collider constraints~\cite{Huitu:2018gbc,Cline:2019okt,Arina:2019tib}, indirect detection searches~\cite{Alanne:2018zjm,Cline:2019okt,Arina:2019tib}, and gravitational wave constraints~\cite{Kannike:2019wsn,Alanne:2020jwx,Zhang:2021alu}. Pseudo-Nambu-Goldstone particle as dark matter can also originate from composite scalars~\cite{Frigerio:2012uc} or directly from the break of UV complete structures~\cite{Ahmed:2020hiw}, however these scenarios do not possess the cancellation mechanism. ~\cite{Ruhdorfer:2019utl} has discussed a possible connection between PNGDM and the composite scenario. 

The assumption of zero momentum transfer limit is commonly adopted in DM direct detection experiments, while it is valid only when all mediator masses are much larger than the momentum transfer, and therefore there exist parameter spaces where the mass of one mediator is comparable to the momentum transfer. Some studies on the direct detection of PNGDM have neglected these parameter spaces and mainly focused on loop corrections~\cite{Azevedo:2018exj,Ishiwata:2018sdi,Glaus:2020ihj,Alanne:2018zjm,Arina:2019tib}. We realized that even on the tree level, these new parameter space can still be approached by current DM direct detection experiments. 
	 
Recent researches~\cite{Baryakhtar:2017dbj,Raj:2017wrv,Bell2018} have pointed out that old neutron star (NS) can capture DM and get reheated, thus giving constraints on DM-neutron interaction by the observation of NS reheating temperature. This mechanism does not require a certain energy threshold, compared with the cases in DM direct detection experiments, for DM-neutron scattering. On the other hand, DM can be accelerated to the relativistic velocity by the strong gravity of NS, and thus the DM-neutron scattering rate will be boosted. These advantages have drawn much attention and NS constraints have been applied to different models~\cite{Baryakhtar:2017dbj,Raj:2017wrv,Bell2018,Chen:2018ohx,Camargo:2019wou,Joglekar:2020liw,Keung:2020teb,Maity:2021fxw,Acevedo:2019agu,Joglekar:2019vzy,Garani:2020wge}. Due to the momentum dependency, the PNGDM benefits more from the DM velocity boosting, resulting in stronger constraints from NS reheating temperature than some other DM candidates.

This paper is organized as follows: In Sec.~\ref{sec:PNGmodel} we take one simple PNGDM model as an example and emphasize its non-zero momentum transfer parameter space; Sec.~\ref{sec:DD} and Sec.~\ref{sec:NS} review the mechanisms of DM direct detection and NS reheating, respectively; in Sec.~\ref{sec:res} we present and discuss the results of new constraints from direct detection experiments and the NS reheating temperature on the PNGDM; and we conclude in Sec.~\ref{sec:con}.

\section{Pseudo-Nambu-Goldstone dark matter model}%
\label{sec:PNGmodel}
The PNGDM models all possess some symmetries that will be broken softly, and the scattering between DM and nucleon will be cancelled by mediators in the zero momentum transfer limit. Here we review the cancellation mechanism with the model mentioned in ~\cite{Barger:2008jx,Gross2017}. The model extends Standard Model by a complex scalar with a global $\mathrm{U}(1)$ symmetry. Its real part will mix with the Higgs boson and the imaginary part will obtain mass through a soft breaking term of the global $\mathrm{U}(1)$ symmetry. The potential of scalars in this model can be written as
\begin{eqnarray}
	V= &&-\mu^2|H|^2-\mu_{S}^2|S|^2+\lambda|H|^{4}+\lambda_{S}|S|^{4}+2\lambda_{SH}|H|^2|S|^2\nonumber\\
	   &&-\frac{{\mu_{S}^{\prime}}^2}{4}S^2+\mathrm{h.c.}\label{potential}
\end{eqnarray}
where $-\frac{{\mu_{S}^{\prime}}^2}{4}S^2$ is the soft breaking term and $\mu_{S}^{\prime}$ can be taken as real without loss of generality. After $H$ and $S$ get their vacuum expectation values, by substituting $H=(0,\frac{v_h+h}{\sqrt{2} })^{T},\ S=\frac{v_{s}+s+i\chi}{\sqrt{2} }$ into Eq.~\eqref{potential}, we obtain the mass term
\begin{eqnarray}
     &&\frac{1}{2}\begin{pmatrix}
	 h&s
     \end{pmatrix}O O^{T}
     \begin{pmatrix}
     2\lambda v_h^2&2\lambda_{SH}v_hv_{s}\\
     2\lambda_{SH}v_hv_{s}&2\lambda_{S}v_{s}^2
 \end{pmatrix}O O^{T}\begin{pmatrix}
     h\\
     s
 \end{pmatrix}\nonumber\\
     =&&\frac{1}{2}\begin{pmatrix}
	 h_1&h_2
     \end{pmatrix}
     \begin{pmatrix}
	 m_1^2&0\\
	 0&m_2^2
 \end{pmatrix}\begin{pmatrix}
     h_1\\
     h_2
 \end{pmatrix},\ \frac{1}{2}{\mu^{\prime}_{S}}^2\chi^2, \label{massterm}
\end{eqnarray}
where $h_1$ and $h_2$ are the mass eigenstates of $h$ and $s$, with mass of $m_1$ and $m_2$, respectively. The orthogonal matrix $O$ transforms $h$ and $s$ to their mass eigenstates. 
Here we choose $h_1$ as the observed Higgs boson, and $m_1$ as the Higgs boson mass. From Eq.~\eqref{massterm} it is clear that the soft breaking term makes $\chi$ a Pseudo-Nambu-Goldstone particle, and thus a DM candidate; otherwise $\chi$ becomes a massless Nambu-Goldstone particle as the soft breaking term goes to zero. The interaction between DM and quark is mediated by $h_1$ and $h_2$, which will cancel out in the zero momentum transfer limit. The coupling terms involved in the interaction are
\begin{eqnarray}
     \frac{m_{f}}{v_h}h\bar{f}f=\frac{m_{f}}{v_h}(h_1O_{11}+h_2O_{12})\bar{f}f,\nonumber\\ 
 \begin{pmatrix}
     h&s
 \end{pmatrix}\begin{pmatrix}
     \lambda_{HS}v_h\\
     \lambda_{S}v_{s}
 \end{pmatrix}\chi^2=\frac{1}{2v_{s}}\begin{pmatrix}
     h_1&h_2
 \end{pmatrix}\begin{pmatrix}
     O_{21}m_1^2\\
     O_{22}m_2^2
 \end{pmatrix}\chi^2.
\end{eqnarray}
Therefore the DM-fermion scattering amplitude is
\begin{eqnarray}
    i\mathcal{M}&&=\frac{m_{f}}{v_hv_{s}}\left(\frac{O_{11}O_{21}m_1^2}{t-m_1^2}+\frac{O_{12}O_{22}m_2^2}{t-m_2^2}\right)\bar{u}(p_3)u(p_1)\nonumber\\
      &&=\frac{m_{f}}{v_hv_{s}}\frac{tO_{12}O_{22}(m_2^2-m_1^2)}{(t-m_1^2)(t-m_2^2)}\bar{u}(p_3)u(p_1)\label{PNGamp} ,
\end{eqnarray}
where we have used the orthogonality of $O$. The amplitude goes to zero in the zero momentum transfer limit, which is usually adopted in the effective field frame where the masses of mediators are considered very large. However, $t$ can not be neglected if one of the mediator mass is comparable or smaller than the momentum transfer $q=\sqrt{-t} =\sqrt{2m_{f}E_{R}} $. In liquid xenon DM direct detection experiments with the normal nuclear recoil energy threshold of about $5~\mathrm{keV}$, the momentum transfer is about $35~\mathrm{MeV}$. Consequently, we can expect that there is no momentum suppress in parameter spaces with $m_2$ of $\sim O(10)~\mathrm{MeV}$.

\section{DM direct detection}%
\label{sec:DD}
To study the direct detection of the PNGDM, we first write out its differential scattering cross section according to Eq.~\eqref{PNGamp} 
\begin{eqnarray}
    \frac{d\sigma_f}{dE_{r}}=\frac{1}{E_{r}^{max}}\frac{1}{16\pi s}\frac{2m_{f}^2}{v_h^2v_{s}^2}\frac{t^2(m_2^2-m_1^2)^2\sin^2 \theta\cos^2 \theta}{(t-m_1^2)^2(t-m_2^2)^2}(2m_{f}^2-\frac{t}{2})\label{pngdcs},
\end{eqnarray}
where $E_{r}$ is the recoil energy of fermion $f$, and we have adopted $O_{12}$ and $O_{22}$ as $\sin \theta$ and $\cos \theta$ respectively. Then we transform the DM-quark interaction to DM-nucleon interaction, in order to compare with experiment results. For scalar interaction with the DM-quark coupling constant $\lambda_{\chi q}$, the DM-nucleon coupling constant can be written as
\begin{eqnarray}
    f_{l\in\{p,n\}}=m_{l}\left(  \sum\limits_{q=u, d, s}f_{q}^{l}\frac{\lambda_{\chi q}}{m_{q}}+\frac{2}{27}f_{G}^{l} \sum\limits_{q=c, b, t}\frac{\lambda_{\chi q}}{m_{q}}\right) ,
\end{eqnarray}
where $l$ represents nucleon, and $m_{l}$ and $m_{q}$ are the mass of nucleon and quark, respectively. $f_{q}^{l}$ and $f_{G}^{l}$ are the corresponding form factors, and can be expressed as, 
\begin{eqnarray}
    f_{u}^{p}=0.026,\ f_{d}^{p}=0.038,\ \\
    f_{u}^{n}=0.018,\ f_{d}^{n}=0.056,\ \\
    f_{s}^{p}=f_{s}^{n}=0.044,\ f_{G}^{l}=1-\sum\limits_{q=u, d, s}f_{q}^{l}
\end{eqnarray}
in the zero momentum transfer limit~\cite{DelNobile:2021icc}. When the momentum transfer can not be neglected, a dipole form factor can be added to describe the momentum suppress~\cite{DelNobile:2021icc,Bringmann:2018cvk,Bell:2020obw,Ye:2017gyb}
\begin{eqnarray}
    f_{q}^{l}(t)=\frac{f_{q}^{l}}{(1-t/Q_0^2)^2},\ f_{G}^{l}(t)=\frac{f_{G}^{l}}{(1-t/Q_0^2)^2}\label{dpformfactor},
\end{eqnarray}
where $Q_0^2=0.71~\mathrm{GeV}^2$. The dipole form factor is important to momentum dependent cross section in the NS constraint and takes part in the total form factor in the direct detection. Further, we can transform the DM-nucleon couplings into DM-nucleus couplings as
\begin{eqnarray}
    \lambda_{\chi N}=Zf_{p}+(A-Z)f_{n}\label{couplings},
\end{eqnarray}
where $Z$ and $A$ are the number of proton and nucleon in the nucleus, respectively. A form factor in nuclear level is necessary since the nucleus can not be taken as an point in large momentum transfer~\cite{DelNobile:2021icc}
\begin{eqnarray}
    \frac{d\sigma_N}{dE_{R}}=\frac{d\sigma_N}{dE_{R}}\bigg|_{\text{PLN}}F^2(E_{R}) ,
\end{eqnarray}
where $E_R$ is nucleus recoil energy and $\frac{d\sigma_N}{dE_{R}}\bigg|_{\text{PLN}}$ is point like nucleus differential cross section. By taking the point like limit, the difference between the DM-quark scattering cross section and the DM-nucleus scattering cross section comes from the corresponding couplings and masses. Since the couplings in the nucleus level links to the quark level by Eq.~\eqref{couplings}, we can write the point like nucleus differential cross section from Eq.~\eqref{pngdcs} as
\begin{eqnarray}
    \frac{d\sigma_N}{dE_{R}}\bigg|_{\text{PLN}}=&&\frac{1}{E_{R}^{max}}\frac{1}{16\pi s}\frac{2}{v_h^2v_{s}^2}\frac{t^2(m_2^2-m_1^2)^2\sin^2 \theta\cos^2 \theta}{(t-m_1^2)^2(t-m_2^2)^2}(2m_{N}^2-\frac{t}{2})\nonumber\\
    &&\times\left(Zm_p\left(  \sum\limits_{q=u, d, s}f_{q}^{p}+\frac{2}{9}f_{G}^{p} \right)+(A-Z)m_n\left(  \sum\limits_{q=u, d, s}f_{q}^{n}+\frac{2}{9}f_{G}^{n} \right)\right)^2,
\end{eqnarray}
where $m_p,\ m_n,\ m_N$ are the mass of proton, neutron, and nucleus respectively.
For direct detection we adopt the total momentum dependent form factor of $^{132}$Xe ~\cite{Fitzpatrick:2012ix,Bramante:2016rdh}
\begin{eqnarray}
    F^2(E_{R})=\frac{e^{-u}}{A^2}\left( A+\sum\limits_{n=1}^{5} c_{n}u^{n}\right)^2,\ \text{with}\ u=\frac{m_{N}E_{R}}{m_{n0}(45A^{-\frac{1}{3}}-25A^{-\frac{2}{3}})~\mathrm{MeV}},
\end{eqnarray}
where $m_{n0}$ is the unit nucleon mass. And $c_1=-132.841,\ c_2=38.4859,\ c_3=-4.08455,\ c_4=0.153298,\ c_5=-0.0013897$~\cite{Vietze:2014vsa}. Finally, the scattering event rate can be written as
\begin{eqnarray}
    \frac{dR}{dE_{R}}=N_{T}\int \frac{\rho_{\chi}v}{m_{\chi}}\frac{d\sigma_N}{dE_{R}}f(\vec{v},\vec{v}_{e}  ) d\vec{v}, 
\end{eqnarray}
where $N_{T}$ is the number of target nucleus, $m_{\chi}$ is the DM mass, and $\rho_{\chi}=0.3~\mathrm{GeV}/\mathrm{cm}^3$~\cite{ParticleDataGroup:2020ssz} is the local DM density. $f(\vec{v},\vec{v}_{e}  )$ is the local DM distribution viewed at Earth
\begin{eqnarray}
    f(\vec{v} , \vec{v}_{e} )=\frac{1}{N_0}e^{-\frac{(\vec{v}+\vec{v}_{e}  )^2}{v_{0}^2}},
\end{eqnarray}
where $N_0$ normalizes the integral $\int f(\vec{v},\vec{v}_{e}) d\vec{v} = 1 $, and $v_{e}$ is the velocity of Earth, which can be approximated as the velocity of the Sun $v_{\odot}=232\text{km/s}$~\cite{Schoenrich:2009bx,Bramante:2016rdh,DelNobile:2021icc}. 
We adopted the same parameter values and followed the detailed calculation as in ~\cite{Bramante:2016rdh,Green:2017odb,Bovy:2012ba,Piffl:2013mla,Lewin:1995rx}. Finally, by comparing the event rate calculated above and that derived from the DM-nucleon scattering cross section upper limits given by direct detection experiments, we can give constraints to the PNGDM model. 

\section{Neutron star reheating temperature}%
\label{sec:NS}
DM will be accelerated to relativistic velocity by the strong gravity from the NS, resulting in an enhanced DM-neutron scatter rate, especially for those whose amplitude is proportional to the momentum transfer. To study the capture of DM by NS, a threshold cross section can be defined as follows: with a DM-neutron scattering cross section $\sigma$ larger than the threshold cross section $\sigma_{t}$, all passing-by DM will be captured by the NS. Then, the capture rate can be written as $r_c=\text{min}(\sigma/\sigma_{t},1)$. 
When determining the form of threshold cross section, if DM mass $m_{\chi}<1~\mathrm{GeV}$, the NS Fermion momentum should be taken into consideration since the momentum transfer might not be enough for the scattering; if DM mass $m_{\chi}>10^{6}~\mathrm{GeV}$, multiple scatterings are required for NS to capture DM. The expressions of threshold cross section for these two cases can be found in ~\cite{Baryakhtar:2017dbj,Raj:2017wrv}, and here we only consider the case of $1~\mathrm{GeV}\le m_{\chi}\le 10^{6}~\mathrm{GeV}$, where a single DM-neutron scattering is enough for NS to capture DM.
For a typical NS with mass $M=1.5M_{\odot}$ and radius $R=10~\mathrm{km}$~\cite{Baryakhtar:2017dbj,Raj:2017wrv,Bell2018}, the threshold cross section can be expressed as $\sigma_{t}=\pi R^2m_{n}/M\approx1.76*10^{-45} \text{cm}^2$, proportional to the NS’s geometric cross section and in the scale of DM direct detection upper limits. 

The impact parameter b for DM at infinity with velocity $v_{\chi}$ can be written as~\cite{Goldman:1989nd}
\begin{eqnarray}
    b=R\frac{v_{es}}{v_{\chi}}(1-2GM/R)^{-\frac{1}{2}},
\end{eqnarray}
where $G$ is the gravitational constant and $v_{es}=(2GM/R)^{1/2}\approx 0.67c$ is the escape velocity of NS. Then the flux of DM scattering with NS can be expressed as
\begin{eqnarray}
    \dot{n}=\pi b^2\frac{\rho_{\chi}}{m_{\chi}}v_{\chi},
\end{eqnarray}
where $n$ is the DM number density. For DM captured by NS, except for those happen to be orbiting the NS, it will scatter with the NS repeatedly. So it is reasonable to assume that all the kinetic energy of captured DM will be finally transferred to NS. The transferred energy from one DM particle can be expressed as 
\begin{eqnarray}
    E_{t}=m_{\chi}(\frac{1}{\sqrt{1-\omega^2} }-1),
\end{eqnarray}
where $\omega=\sqrt{v_{\chi}^2+v_{es}^2} $ is the DM velocity when it reaches the NS. Therefore, the absorption energy flux of NS can be expressed as
\begin{eqnarray}
    \dot{E}=\pi b^2\frac{\rho_{\chi}}{m_{\chi}}E_{t}v_{\chi}.
\end{eqnarray}
Supposing there are no other sources to reheat an old NS, then when in equilibrium the NS reheating temperature if measured at infinity can be expressed as
\begin{eqnarray}
    T&&=\left(\frac{\int  f(\vec{v}_{\chi} , \vec{v}_{N} ) \pi b^2\frac{\rho_{\chi}}{m_{\chi}}E_{t}v_{\chi}r_{c}d\vec{v}_{\chi}}{4\pi\sigma_{SB}R^2}\right)^{\frac{1}{4}}(1-2GM/R)^{\frac{1}{2}}\\
     &&=\left(\int  f(\vec{v}_{\chi} , \vec{v}_{N} )r_{c}\frac{ \rho_{\chi}E_{t}v_{es}^2}{4m_{\chi}\sigma_{SB}v_{\chi}}(1-2GM/R)d\vec{v}_{\chi}\right)^{\frac{1}{4}},
\end{eqnarray}
where $\sigma_{SB}$ is the Stefan-Boltzmann constant and $\vec{v}_{N}$ is the velocity of NS, which is assumed to be the same as the velocity of the Sun. $f(\vec{v}_{\chi} , \vec{v}_{N} )$ is the velocity distribution of DM in the NS frame derived from the Standard Halo Model in the galaxy.
The dipole form factor in Eq.~\eqref{dpformfactor} has been taken into account and implemented in $r_c$. In this work $\rho_{\chi}=0.3~\mathrm{GeV}/\mathrm{cm}^{3}$ is used for both direct detection and NS temperature calculation, resulting in the maximal NS reheating temperature of $\sim1564~\mathrm{K}$.  
~\cite{Baryakhtar:2017dbj} calculated the maximal temperature
\begin{eqnarray}
    T_{max}=\left( \frac{ \rho_{\chi}E_{t}v_{es}^2}{4m_{\chi}\sigma_{SB}\bar{v}_{\chi}}(1-2GM/R)\right)^{\frac{1}{4}}
\end{eqnarray}
to be about $1750~\mathrm{K}$ with $\bar{v}_{\chi}$ being the mean velocity of DM. Furthermore, ~\cite{Bell2018} considered the distribution of $\frac{1}{v_{\chi}}$ and resulted in a $T_{max}$ of $1700~\mathrm{K}$. Here we have also taken into account the velocity-dependent $r_{c}$ and $E_{t}$. 
Since DM is non-relativistic in the galaxy, its velocity is actually much smaller than the escape velocity of NS, thus has very little influence to $r_{c}$ and $E_{t}$. Nevertheless, in this work we use the full integral of velocity distribution to calculate the temperature. 

After the potential discovery of nearby old NS by FAST~\cite{Nan:2011um,Han:2021ekd}, future telescopes might be able to measure the temperature of the NS whose temperature is lower than $1000~\mathrm{K}$~\cite{Baryakhtar:2017dbj,Raj:2017wrv} and give constraints to the DM-nucleon interaction. The temperature contours used in the next section represent the sensitivity of future NS temperature measurements.

\section{results and discussion}%
\label{sec:res}
We have chosen two scenarios of parameters, as listed in Table~\ref{tab:paras}, to constrain PNGDM model from the direct detection and NS temperature, with the Higgs boson mass $m_1=125.10~\mathrm{GeV}$ and the vacuum expectation $v_h=246.22~\mathrm{GeV}$ taken from~\cite{ParticleDataGroup:2020ssz}.
\begin{table}[ht]
 \centering
\begin{tabular}{c c c c c }
 \hline
  &$m_2$&$m_{\chi}$&$\lambda_{S}$  &$\lambda_{SH}$   \\
 \hline
    PNG1&variable&variable&0.1&0.01\\
    PNG2&1~$\mathrm{GeV}$&variable&0.1&variable\\
 \hline
\end{tabular}
\caption{Two scenarios of parameters used to constrain PNGDM model.}
    \label{tab:paras}
\end{table}
Other parameters not specified can be calculated accordingly. Variables in scenario PNG1 are chosen to be $m_2$ and $m_{\chi}$, while in scenario PNG2 they are $m_{\chi}$ and $\lambda_{SH}$. Variable $m_2$ determines the area where the momentum transfer can not be neglected, and variable $\lambda_{SH}$ represents the coupling strength between $h$ and $s$. These two settings will give insights to the internal structure of the PNGDM model.    
We use the recent PandaX-4T experiment results~\cite{PandaX:2021osp} and a few NS reheating temperature to give constrain to the PNGDM model, together with the DM relic abundance constraints from the Planck experiment~\cite{Planck:2018vyg} and the indirect detection constraints from the AMS experiment~\cite{AMS:2014bun} and the Fermi-LAT experiment~\cite{Fermi-LAT:2015att}.
The DM relic abundance $\Omega_\chi h^2$ is calculated with \texttt{MadGraph}~\cite{Alwall:2014hca} plugin \texttt{MadDM}~\cite{Ambrogi:2018jqj}. The model imported by \texttt{MadDM} is implemented in \texttt{FeynRules~2}~\cite{Alloul:2013bka}. We also use \texttt{MadDM} to calculate the DM annihilation cross section for indirect detection constraints.

\begin{figure}[ht]
    \centering
    \subfigure{\includegraphics[width=0.49\textwidth]{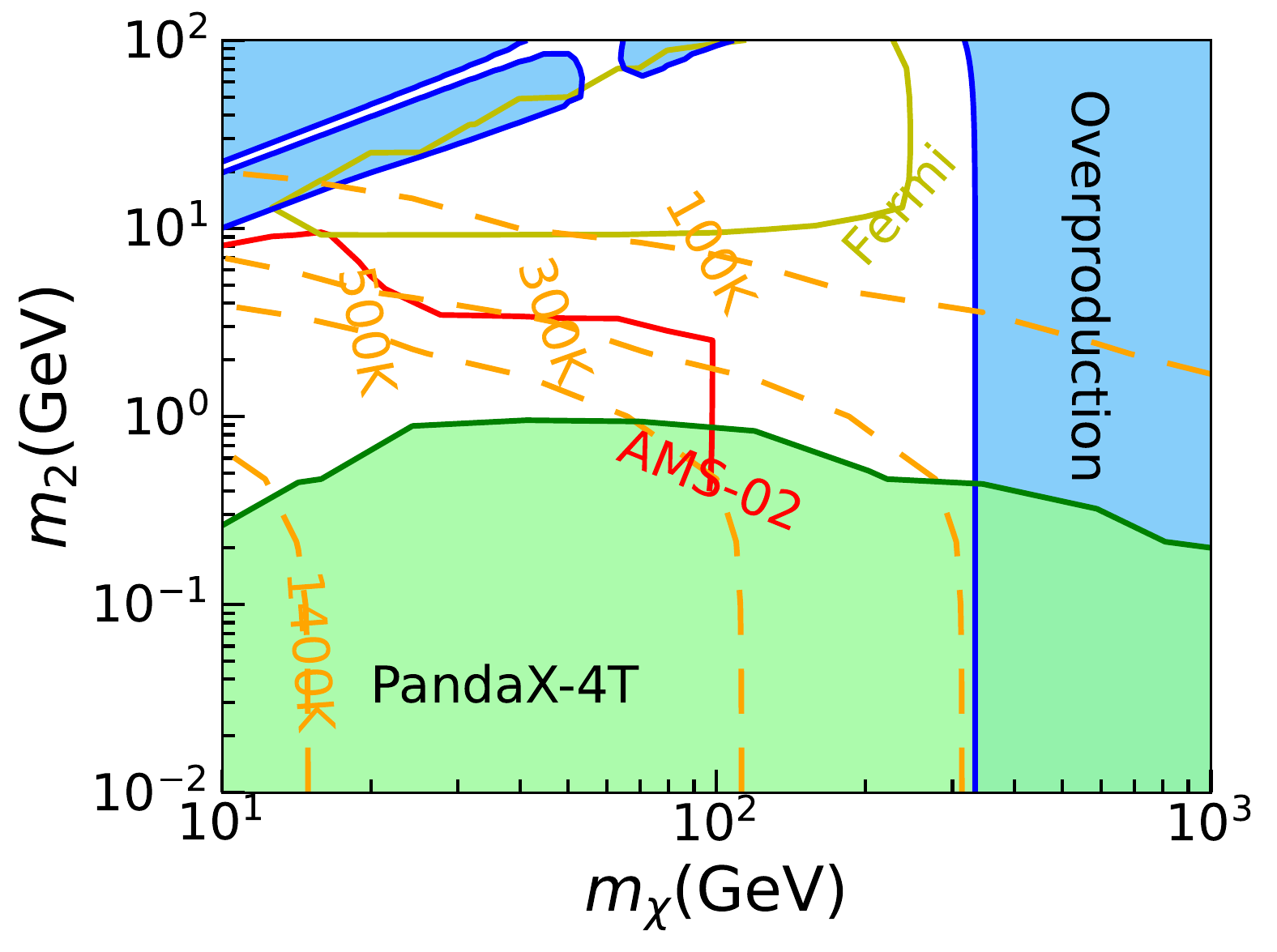}}
    \subfigure{\includegraphics[width=0.49\textwidth]{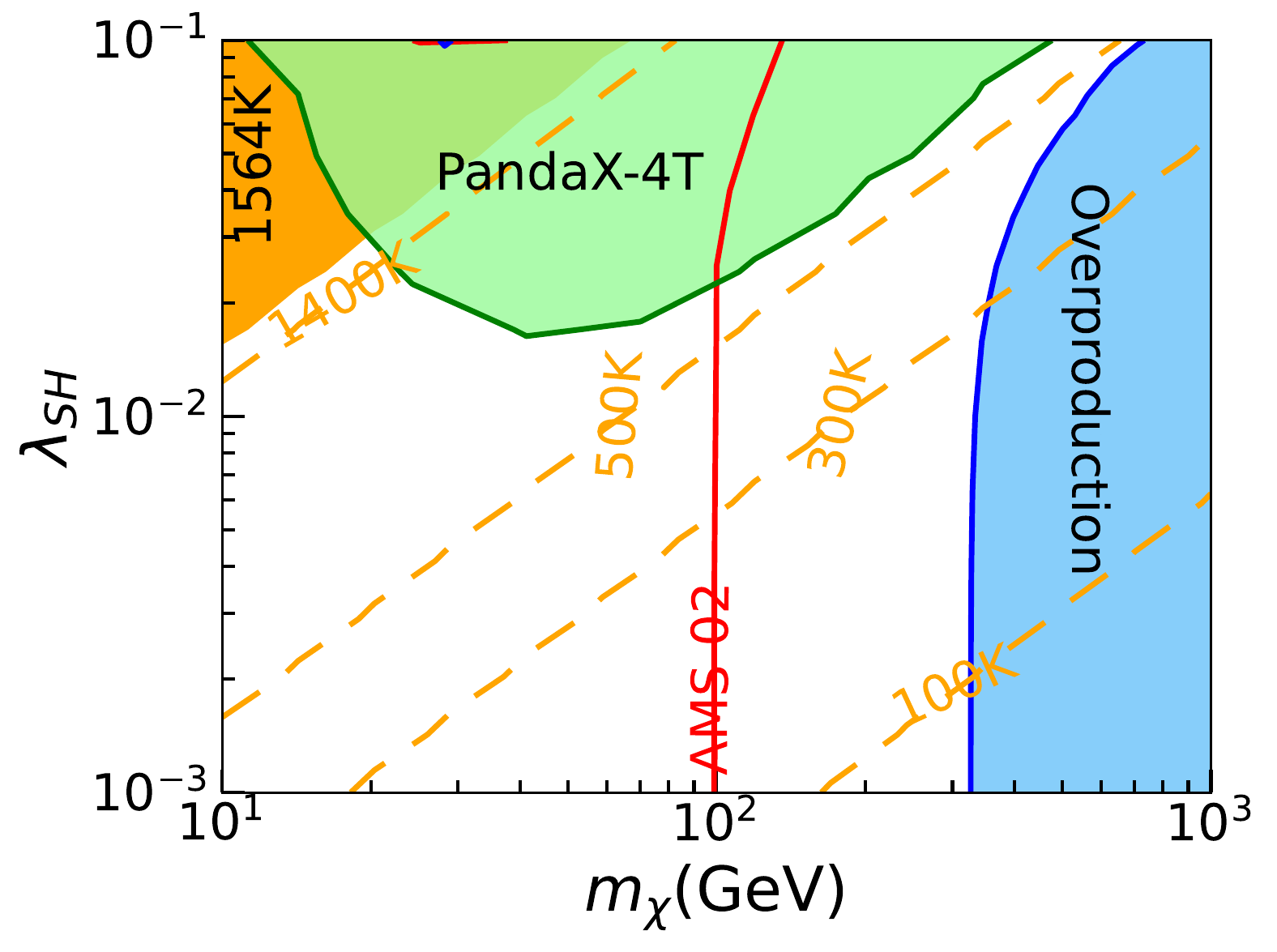}}
    \caption{Constraints on the PNGDM model from the direct detection and NS temperature for two scenarios PNG1 and PNG2 are shown in the left and right panels, respectively. Green areas are excluded by the PandaX-4T experiment~\cite{PandaX:2021osp}. Orange areas are parameters space corresponding to the maximal NS reheating temperature of $1564~\mathrm{K}$, while reheating temperatures of $100~\mathrm{K}$, $300~\mathrm{K}$, $500~\mathrm{K}$ and $1400~\mathrm{K}$ are labeled by dashed orange lines, respectively. Light blue areas are excluded by the Planck experiment~\cite{Planck:2018vyg}. The observed DM relic abundance lies in the blue lines. The red line is derived from the AMS experiment results ~\cite{AMS:2014bun}, of which the left area is excluded. The yellow contour represents constraints by the Fermi-LAT experiment~\cite{Fermi-LAT:2015att} with the inside area excluded.}
    \label{fig:res}
\end{figure}

Fig.~\ref{fig:res} shows the constraints on the PNGDM model from the direct detection and NS temperature, with the left and right panels corresponding to the two scenarios PNG1 and PNG2, respectively. The green areas are excluded by the PandaX-4T experiment at 90\% confidence level~\cite{PandaX:2021osp}, assuming a total detection efficiency of 78\% in average. In the left panel, the excluded area lies in the middle where $m_2$ is about $O(10)~\mathrm{MeV}$ as expected, and $m_2$ can be as large as almost $1~\mathrm{GeV}$ with $\lambda_{SH}=0.01$. As one can see in the right panel, the constraints grow as $\lambda_{SH}$ increases. The light blue areas are excluded by the Planck experiment~\cite{PandaX:2021osp}.
The observed relic abundance $\Omega_\chi h^2=0.12\pm 0.0012$ is shown as the blue lines. 
A band structure appears in the light blue area on the top-left of the left panel, caused by the resonant annihilation when the DM mass is around half of $m_1$ and $m_2$. Also, when $m_2$ is smaller than the mass of Higgs boson, DM can not be too heavy, either. It is clear that in most parameters space excluded by the current direct detection, the DM relic abundance is less than observation, meaning that in these areas the PNGDM only contributes parts of DM constituents. PNGDM with mass of about $330~\mathrm{GeV}$ can meet the correct relic abundance, while the corresponding $m_2$ can be excluded by the direct detection to about $0.43~\mathrm{GeV}$. It's worthwhile to note that a point in the top-middle of the right panel meets the correct DM relic abundance.
Red line and yellow contour represent constraints from indirect detection, which are obtained by comparing the calculated annihilation cross section with the model independent bounds given by Fig.~11 of ~\cite{Elor:2015bho}. The left area of the red line labelled as ``AMS-02'' is excluded by the AMS experiment~\cite{AMS:2014bun} results for $\mu^{+}\mu^{-}$ final states, and the area inside the yellow contour labelled ``Fermi'' is excluded by the Fermi-LAT experiment~\cite{Fermi-LAT:2015att} results for $b\bar{b}$ final states. 
In our parameter settings when $m_2< m_{\chi}$ DM mainly annihilates into two $h_2$, and $h_2$ will mainly decay into two $s$ quarks and two muons in the PNG2 scenario, while in the PNG1 scenario $h_2$ will mainly decay into $e^{+}e^{-}$, $\mu^{+}\mu^{-}$, $\tau^{+}\tau^{-}$ and $b\bar{b}$, respectively, as $m_2$ increases. Here we adopt the constraints for $\mu^{+}\mu^{-}$ and $b\bar{b}$ final states for illustration purpose.

On the other hand, the orange areas are parameters space where all DM transiting the NS will be captured, leading to the maximal reheating temperature of about $1564~\mathrm{K}$. The DM-nucleon cross section in these areas is larger than the threshold cross section $\sigma_{t}$, which is in the order of $10^{-45}~\text{cm}^2$. Therefore, parts of these parameters space have been excluded by the current direct detection, and the momentum enhancement effect plays an important role in the rest area. The dashed orange lines correspond to the NS temperature sensitivities to the PNGDM model, with temperature of $100~\mathrm{K}$, $300~\mathrm{K}$, $500~\mathrm{K}$ and $1400~\mathrm{K}$, respectively. The NS can cool down to $100~\mathrm{K}$ after $10^{9}$ years~\cite{Yakovlev:2004iq}. If current and future observations could find an old nearby NS with temperature of $100~\mathrm{K}$ in the future, the parameters space on the left of the $100~\mathrm{K}$ temperature line in Fig.~\ref{fig:res} could be excluded, since in these areas DM can solely reheat the NS to a temperature larger than $100~\mathrm{K}$.
One may also note that the rotochemical heating effect can heat NS to temperature much higher than $O(1000)~\mathrm{K}$ and thus might conceal the DM heating effect. ~\cite{Hamaguchi:2019oev} has studied the NS temperature with both rotochemical and DM heating effects taken into consideration, and concluded that the rotochemical heating is ineffective if the initial period of NS is relatively large. An observation of NS with temperature lower than the maximal reheating temperature induced by DM can give constraints to DM models.

\section{Conclusion}
\label{sec:con}
In this work we have introduced two new phenomenology constraints to the PNGDM model: the direct detection constraints from PandaX-4T, and future NS temperature observations. 
The current direct detection can exclude large parameters space of the PNGDM model, while the NS temperature observations have very good sensitivities.
These two phenomenology constraints can be applied to different PNGDM models in the same paradigm. 
Besides, the direct detection of non-zero momentum transfer parameters space is worth exploring for other momentum-suppressing models.
The effects of full integral of velocity distribution in NS constraint are expected to be significant in dealing with relativistic DM.  

\section*{Acknowledgements}
This work was supported in part by grants from the Ministry of Science and Technology of China (No. 2016YFA0400301), and from National Natural Science Foundation of China (No. 12090062).

\section*{Note added}
During the review of this manuscript, we noticed that ~\cite{Abe:2021vat} had been posted. ~\cite{Abe:2021vat} discusses several phenomenology constraints on the PNGDM model, including those from the direct detection which is similar as what is discussed in this manuscript, but in different scenarios.

\section*{References}

\bibliography{ref}




\vspace{-.3cm}

\end{document}